# A Synchronous, Reservation Based Medium Access Control Protocol for Multihop Wireless Networks

Jennifer C. Fang
Dept. of Electrical and Computer Engineering
University of California, San Diego
La Jolla, CA
jenny@alumni.caltech.edu

George D. Kondylis
Home and Wireless Networking Business Unit
Broadcom Corporation
Sunnyvale, CA
kondylis@broadcom.com

*Abstract—* We describe a new synchronous and distributed medium access control (MAC) protocol for multihop wireless networks that provides bandwidth guarantees to unicast connections. Our MAC protocol is based on a slotted time division multiple access (TDMA) architecture, with a multi-mini-slotted signaling phase scheduling data transmissions over slots in the following data phase. Resolving contentions at the beginning of a frame allows for effective utilization of bandwidth. Our protocol essentially combines the benefits of TDMA architecture with the distributed reservation mechanism of IEEE 802.11 MAC protocol, thereby performing well even at high loads. We implement a two-way handshake before each data slot to avoid deadlocks, a phenomena that plagues 802.11. Through theoretical analysis, we derive the system throughput achieved by our MAC protocol. We implemented our MAC protocol into *ns-2* simulator, and demonstrate its vast superiority to IEEE 802.11 and a synchronous MAC protocol CATA through extensive simulations.

## I. INTRODUCTION

Research interest in multihop radio networks has accelerated in recent years, with the widespread deployment and the use of wireless LANs worldwide. The underlying MAC protocol in existing WLANs is IEEE 802.11 MAC DCF [1] , which we shall refer to as MAC-802. This protocol though asynchronous and distributed in nature, performs poorly at high loads. It is well known that the scheduling transmissions in TDMA sequence greatly increases capacity [2]. However, such scheduling requires sharing of global topology information among nodes, making it unsuitable for multihop networks. Our synchronous MAC protocol, which is henceforth referred to as MAC-RSV, combines the desirable features of slotted architecture with the distributed bandwidth reservation of MAC-802. In stark contrast to existing MAC protocols, MAC-RSV is specifically designed to support bandwidth reservations for real-time traffic. A notable feature of MAC-RSV is its immunity to the infamous hidden terminal problem that plagues MAC-802.

MAC-RSV is synchronous, based on slotted time, and supports unicast transmissions. Nodes are assumed to be synchronized with reference to a global clock, as required by any TDMA-based protocols. This synchronization can be achieved by either using a GPS at each node, or through periodic beacons from the access point as done by IEEE 802.11 PCF. The signaling section consists of a series of three mini-slot reservation sequences and precedes the data section (which consists of multiple data slots). The three mini-slot reservation sequences are used by the nodes to contend for data slot reservations. Nodes can reserve multiple data slots in a frame whenever they win a contention. Complete separation of the signaling section from the data section provides two advantages: the fraction of signaling bandwidth can be optimized given a specific data bandwidth; and the congestion during new reservations becomes independent of the existing load in the network.

Another unique feature of our protocol is the inclusion of an additional mini-slot preceding each data slot. Intended receivers transmit a short beacon during this mini-slot to their corresponding transmitters, thereby eliminating collisions caused by *deadlocks* (to be defined in section 2). Furthermore, since these mini-slots reaffirm reservations, they provide flexibility of using large data frames as well as increased robustness in mobile environments.

Previous research efforts on QoS enabled MACs include algorithms on *broadcast scheduling* [8] [7], and some more relevant, *topology dependent* scheduling algorithms [3] [4]. Here we focus specifically on supporting bandwidth reservations for unicast connections. A synchronous MAC protocol CATA [3] uses signaling mini-slots inside each data slot. Terminals that want to contend for a specific data slot can only do so in the corresponding mini-slot. This has a very adverse effect on the throughput, even under moderate load. This is because when the number of available data slots decreases, contentions for these slots increase dramatically. In constrast, MAC-RSV resolves contentions collectively in the multi-mini-slotted signaling phase, and each data slot can be contended in any of the mini-slots. In simulation, our protocol achieves significantly higher throughput (about 200% more than 802.11) and lower delays, as indicated by extensive simulations using ns-2. MAC-RSV also outperforms CATA by more than 55% in throughput.

In the following section, we describe the operation of our MAC-RSV protocol. Section III proves its correctness. We derive the system throughput through theoretical analysis in section IV. Section V presents comparison results of MAC-RSV with 802.11 DCF MAC, using the *ns-2* simulation tool.

## II. PROTOCOL DESCRIPTION

The slot structure is shown in Fig. 1. The frame is subdivided into two sections, one for signaling, during which nodes contend for data slots, and the other for data transmissions. The signaling phase contains a number of *mini-slot triplets*. A node requesting data slots transmits a *request to send* (RTS) in the first mini-slot of a triplet, while nodes that receive this RTS transmit a *(not) clear to send* ((N)CTS) in the second mini-slot of the same triplet. If the sender of RTS receives a CTS, it transmits a *confirm* (CONF) packet in the last slot of the triplet. In addition, each data slot contains a *receive beacon* (RB) mini-slot at the beginning, and an *acknowledgement* (ACK) mini-slot at the end. These mini-slots are used by the active receivers of that data slot.

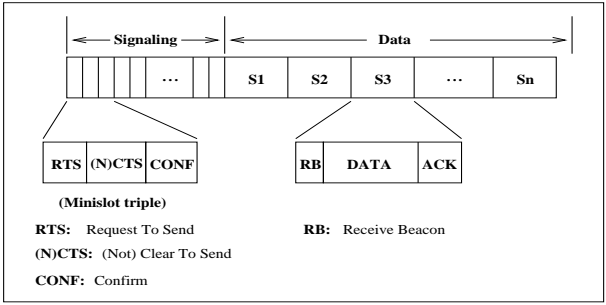

Fig. 1. Slot structure of signaling and data parts of the TDMA frame.

### A. Reservation Mechanism

Each node classifies each *data* slot in the following way:
1) Reserved for transmission (RT).
2) Reserved for reception (RR).
3) Free for transmission (FT). A node can transmit but not receive. A node marks a slot FT if it has not marked it RT or RR, and that a neighbor has marked the same slot RT.
4) Free for reception (FR). A node can receive but not transmit. A node marks a slot FR if it has not marked it RT or RR, and that a neighbor has marked the same slot RR.
5) Free for transmission or reception (FTR). A node has not marked the slot RT or RR, and no neighbor have reserved this slot.

Prior to reservation, a node randomly chooses a data slot from the set of FT and FTR slots. During the signaling section of the frame the node picks a mini-slot and sends a RTS[1], in which it specifies its own ID, the intended receiver's ID and the slot(s) to be reserved. Upon receiving a correct CTS during the paired (N)CTS mini-slot, the reserving node marks the slots as RT and replies with a CONF message. It is possible that the CTS acknowledges only a subset of the requested slots. The node remains silent in the CONF mini-slot if it receives an NCTS or sense a collision in the (N)CTS mini-slot.

---
[1]We will assume a p-persistent policy for contention resolution.

The intended receiver of the RTS packet upon collision-free reception checks if any of the requested slots are among its FR or FTR slots. If so, the receiver issues a CTS in the paired (N)CTS slot and waits for CONF from the transmitter, whereupon it marks the slots RR. The receiver may only acknowledge a subset of the requested slots if it finds only these slots with FR or FTR status. If none of the requested slots are in the FR or FTR sets, the receiver remains silent during the (N)CTS slot.

When a node other than the intended receiver receives a RTS collision-free, it checks if any of the requested slots are in its RR set. If not, the node remains silent in the (N)CTS mini-slot and waits for CONF from the transmitter, whereupon it marks the slots FT. If any of the requested slots are in the node's RR set, in order to protect existing reservations, the node issues an NCTS in order to jam any possible CTS issued by the intended receiver of the RTS. Finally, a node that detects a collision in an RTS mini-slot issues an NCTS to protect its own RR slots which the RTS sender might be requesting from another neighbor.

### B. Receiver Beacon (RB) and Acknowledgment (ACK)

A node that has marked a data slot RR (i.e., active data receiver in this slot) transmits a RB at the beginning of the slot. The beacon contains the ID of the active data transmitter. This short beacon is especially useful in resolving deadlocks[2] and in case of mobility. Correct reception of RB verifies that deadlock has not occurred. Collision in RB signifies deadlock or movement of other receivers in the neighborhood. In this case, the active transmitters opt to contend for new data slots.

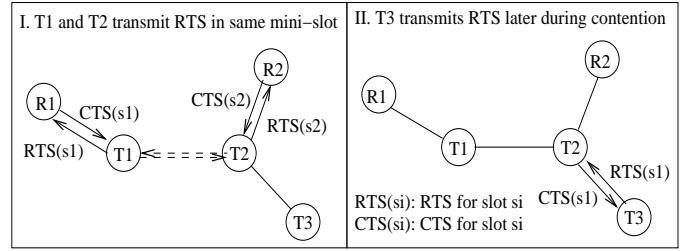

Fig. 2. An example of deadlock during reservation.

Fig. 2 depicts a potential deadlock situation. $T_1$ and $T_2$ send RTS in the same mini-slot. While $T_1$ successfully reserves slot $s_1$, and $T_2$ reserves $s_2$, neither have knowledge of the other's reservation (because neither can hear the other's RTS). Later, $T_3$ is able to successfully reserves $s_1$ at $T_2$, because $T_2$ has not marked $s_1$ FT. A potential collision can occur at $T_2$ in slot $s_1$. However, in the RB mini-slot at the beginning of $s_1$, collision occurs at $T_1$ from the RB's sent by both $R_1$ and $T_2$. Consequently, $T_1$ defers transmission in $s_1$ while $T_3$ transmits successfully.

Additionally, the transmission of RB by a receiver re-affirms the reservation to the neighbors of the receiver, who mark the slot as FR and avoid contending for it. This is especially useful under mobility, when new nodes move to the receiver's vicinity and should be notified that this slot is already reserved for reception by a neighbor. Finally, the absence of RB signifies that the receiver has left the neighborhood.

---
[2]Deadlock occurs when two nodes *without common neighbors* transmit RTS simultaneously and therefore cannot hear each other's transmission.

Receivers notify transmitters of correct data reception by means of an ACK at the end of each data slot. This is particularly vital under mobility and during deep fades.

## III. CORRECTNESS OF THE PROTOCOL

The following theorem proves the correctness of our reservation mechanism.

*Theorem 1:* If any two nodes have at least one common neighbor, then reservations are done correctly, that is when a slot is successfully reserved it is impossible for the data packet to experience collision.

*Proof:*

A slot $s_0$ is successfully reserved when the receiver ($R$) receives a collision-free RTS and a subsequent CONF, while the sender ($S$) receives a collision-free CTS, and the RTS, CTS, and CONF contain $s_0$ in their fields. For $R$ to receive RTS, no other node except $S$ in the neighborhood of $R$ ($N(R)$) sent RTS (including $R$ itself), hence no node in $N(R)$ will reserve any slot for transmission in the current RTS/(N)CTS/CONF mini-slot sequence. For $R$ to send CTS, $s_0$ must be in its FR or FTR slots, and hence no node in $N(R)$ reserved $s_0$ for transmission during past contention mini-slots. For otherwise, $R$ would have certainly heard the relevant RTS and CONF and would have put $s_0$ in FT. For the rest of the contention part of the frame, if $R$ receives an RTS that contains $s_0$, or detects collision of RTSs, it will transmit NCTS that jams any CTS possibly received by the sender of the RTS, thus prohibiting the sender to reserve $s_0$ (or any other slot, for that matter). Therefore, by the end of the contention period $S$ will be the unique transmitter for slot $s_0$ in $N(R)$.

On the other hand, once $R$ has transmitted a CTS, $S$ will receive it collision free in the paired (N)CTS slot, if and only if: (a) $S$'s RTS was received collision-free from all nodes in $N(S)$, for otherwise, some node in $N(S)$ would have transmitted NCTS to jam any possible CTS, and (b) the slots requested in $S$'s RTS were not among the RR slots of any of the receivers of the RTS, for else, again some node in $N(S)$ would have transmitted NCTS to jam $S$'s reception of CTS. Thus, upon reception of CONF, all nodes in $N(S)$, except $R$, will mark the slots contained in the RTS as FT and none will become a receiver for $s_0$ during subsequent contention mini-slots. Therefore, by the end of the contention period, $R$ will be the unique receiver in $N(S)$, for slot $s_0$. ∎

## IV. THROUGHPUT ANALYSIS

We present here a throughput analysis for our protocol for the special case of a fully connected network to reflect the worst-case scenario. Specifically, we find the average utilization of data slots in a TDMA frame. We assume that our TDMA frame has $K$ mini-slot triples and $N$ data slots. Packet arrivals at nodes are Poisson distributed, with mean interarrival time $\tau$. New packet always arrives at a new node (infinite terminal population), and nodes die after packet transmission. The packet size is a truncated geometric random variable $L$ with parameter $q$, $L \leq N$ in the number of slots, i.e., the probability distribution $P[L = l] = \frac{1-q}{1-q^N} q^{l-1}$.

A packet remains in the node's buffer until the node obtains a successful reservation for the whole packet, that is for $L$ data slots. Let $N_c(n)$ denote the number of contending nodes in the beginning of the $n$-th TDMA frame. We assume that new arrivals during a frame start contending from the next frame, therefore $N_c(n)$ is non-increasing during contention. Let $J(n)$ denote the number of nodes that obtain successful reservations for the $n$-th TDMA frame. It is clear that, conditioned on $N_c(n)$, $J(n)$ does not depend on any $J(n-k)$, $k > 0$. The number of contenders evolve as:

$$N_c(n+1) = N_c(n) + \text{new arrivals in } n^{th} \text{ frame} - J(n)$$

Moreover since the arrival process is Poisson and hence memoryless and $J(n)$ conditioned on $N_c(n)$ is idependent of the past, we can see that $N_c(n)$ is a Markov chain. To find the transition probabilities we need to first find the *pdf* of $J(n)$, conditioned on $N_c(n)$, $P[J(n) = j | N_c(n) = n_c]$. To this end, say that up to the beginning of the $k$-th RTS mini-slot $M = m < k$ nodes have already obtained reservations successfully.

Since the network is fully connected, all nodes have knowledge of the number of slots reserved, say $R \leq N$. Let $m$ nodes cease contending, while an additional nodes with packets larger than $N - R$ will not contend, but instead wait for the next TDMA frame. Now, the *pdf* of the number of reserved slots $R$ is given by the convolution of $m$ truncated geometric distributions, conditioned on $R \leq N$:

$$P[R = r] = P[L_1 + \cdots + L_m = r | L_1 + \cdots + L_m \leq N] \quad (1)$$

where $L_i$ is the size of the packet of the $i$-th succeeding node. Since $P[L_1 + L_2 + \cdots + L_m = r]$ ($\equiv P_{mL}(r)$) is the convolution of $m$ truncated geometric distributions can be rewritten as:

$$P[R = r] = q^r \frac{\prod_{i=1}^{m-1}(r-i)}{\sum_{j=m}^{N} q^j \prod_{i=1}^{m-1}(j-i)} \quad (2)$$

Furthermore,

$$P[L > N - r | R = r] = 1 - \frac{1-q^{N-r}}{1-q^N} \equiv s$$

The number of nodes, $N_d$, that will drop out of contention for the current frame due to their packet size, will follow a binomial distribution:

$$P[N_d = n_d | R = r] = \binom{n_c - m}{n_d} s^{n_d}(1-s)^{n_c-m-n_d} \quad (3)$$

Finally, conditioned on the number of previous successes ($M = m$) and the number of contending nodes in the beginning of the contention period ($N_c = n_c$), the probability of a

successful reservation during the $k$-th RTS, $P_{\text{succ}}(m, n_c)$ will be given by:

$$P_{\text{succ}} \equiv P[\text{success in } k\text{-th RTS} | M = m, N_c = n_c] = \ldots$$
$$= \sum_{r=m}^{N} \sum_{n_d=0}^{n_c-m} g(n_c - m - n_d) P[N_d = n_d | R = r] P[R = r] \quad (4)$$

where $g(n) = np(1-p)^{n-1}$ is the probability of a single RTS transmission. For $m = 0$, $P_{\text{succ}}(0, n_c) = g(n_c)$.

Equation (4) implies that the probability of a successful reservation during a mini-slot triple depends on the number of prior reservations during the contention period. Consequently, to calculate the probability distribution of the number of successes, we need to consider all the possible success sequences during the $K$ RTS/CTS/CONF mini-slot triples. Specifically, let $J = j$ successes during the whole contention period, and let $\{h_1, h_2, \ldots, h_j\}$ denote the positions, in ascending order, of successful reservations in the $K$ RTS/CTS/CONF minislots; for example if $K = 4$, $J = 2$ and successful reservations occured in the first and third mini-slot triples, then $h_1 = 1, h_2 = 3$. Then, it is easy to see that $P[J = j | N_c = n_c]$ ($\equiv P_j$) for $j > 0$ will be calculated as follows:

$$P_j = \begin{cases} \sum_{\substack{\text{all } \binom{K}{j} j-\text{tuples} \\ \{h_1, h_2, \ldots, h_j\}}} [1 - P_{\text{succ}}(0, n_c)]^{h_1-1} \cdot P_{\text{succ}}(0, n_c) \\ \times [1 - P_{\text{succ}}(1, n_c)]^{h_2-h_1-1} \cdot P_{\text{succ}}(1, n_c) \\ \times [1 - P_{\text{succ}}(j-1, n_c)]^{h_j-h_{j-1}-1} \cdot \\ \times P_{\text{succ}}(j-1, n_c) \\ \times [1 - P_{\text{succ}}(j, n_c)]^{K-h_j}, \end{cases} \quad (5)$$

Also, $P[J = 0 | N_c = n_c] [1 - P_{\text{succ}}(0, n_c)]^K$. It is worth noting that, if the probability of success in any mini-slot triple did not depend on the number of previous successes (i.e., if $P_{\text{succ}}(m, n_c)$ was independent of $m$), equation (5) would be a simple binomial distribution of $j$ successes out of $K$ trials, with success probability $P_{\text{succ}}(n_c)$ for each trial.

Using equations (2) through (4) we can calculate $P[J = j | N_c = n_c]$ from equation (5). Now we find the transition probabilities $P[N_c(n+1) = j | N_c(n) = i]$, also denoted by $TP$ for the Markov chain $N_c(n)$. Denoting $N_A$ as the number of new arriving packets during the $n$-th TDMA frame, we have

$$TP = \sum_{k=[i-j]^+}^{\min\{i,K\}} P[N_A = j - i + k | N_c(n) = i, J(n) = k] \times \cdot$$
$$\ldots \times P[J(n) = k | N_c(n) = i]$$
$$= e^{-T/\tau} \sum_{k=[i-j]^+}^{\min\{i,K\}} \frac{(T/\tau)^{j-i+k}}{(j-i+k)!} P[J(n) = k | N_c(n) = i] \quad (6)$$

where $[x]^+ = x$, if $x \geq 0$ and zero otherwise. From the transition probabilities for $N_c(n)$ we can calculate the stationary probability distribution $\pi(n_c)$, if one exists. Finally, we calculate the *pdf* for the number of reserved slots in a frame as follows:

$$P(R = r) = \sum_{n_c=1}^{\infty} \sum_{j=1}^{\min\{n_c, K, r\}} P(R = r | J = j) \times \ldots$$
$$\ldots P(J = j | N_c = n_c) \pi(n_c), \ r > 0. \quad (7)$$

TABLE I
SIMULATION PARAMETERS FOR FIG. 4 AND FIG. 5

| Reservation slots (S) | 14 |
|---|---|
| Data Slots (D) | 25 |
| Persistence probability | 0.175 |
| Control frame size | 20 bytes |
| Data frame size | 1044 bytes |

## V. PERFORMANCE EVALUATION

We evaluate the performance of our protocol through simulations, using the network simulator, *ns-2*. We compare the throughput and delay performance of our MAC-RSV protocol against IEEE 802.11 MAC DCF as implemented in *ns-2*. The physical layer parameters of the radios approximate Lucent's WaveLAN DSSS interface, with 250 meters nominal range and bandwidth 2 Mbps.

We consider two simulation scenarios: a static mesh network and a mobile network, each with 25 nodes. In the mesh network, nodes are positioned 200 meters apart as shown in Fig. 3. This is clearly the worse case scenario for deadlocks to occur, since no two nodes have any common neighbors. Nodes in the mobile network are scattered over an $1500m \times 300m$ area. They move at a speed of 20 $m/sec$ according to the random waypoint model. The traffic model at nodes is an independent Poisson process with mean arrival rate $\lambda$. Refer to table I for other simulation parameters. We simulated one-hop communications only, since we are interested in MAC layer statistics.

### A. Simulation Results

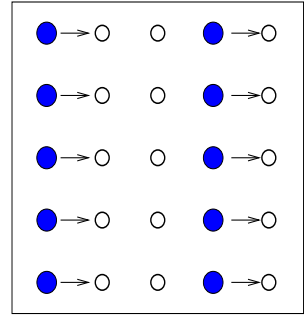

Fig. 3. The 25-node mesh network.

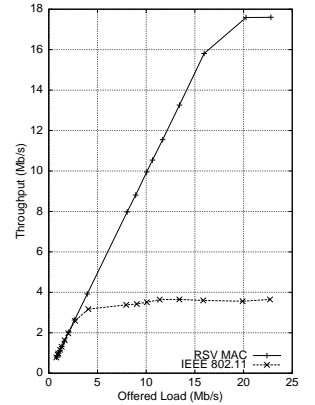

Fig. 4. Throughput for communication scenario shown in Fig. 3

For the aforementioned mesh network structure, Fig. 3 shows a communication scenario that achieves the maximum aggregate throughput, which is 20 Mbps. Shaded nodes can simultaneously transmit without conflict. The throughput achieved by MAC-RSV is almost 18 Mbps, at least **4** times that of 802.11, as shown in Fig. 4. The results show that even under moderate contentions, 802.11 MAC's backoff algorithm defers transmissions and limits throughput severely. On the other hand, contention in our MAC-RSV is much lower, since the nodes that receive a RTS or a CTS not addressed to them for a specific slot can still contend or receive for the same slot, respectively.

Simulation results for the mesh and mobile networks with random communication scenarios indicate similar trends, highlighting the robustness of MAC-RSV in the presence of hidden/exposed terminals. Throughput results for the mesh network are shown in Fig. 5(a). For very low loads, 802.11 slightly outperforms MAC-RSV, due to lack of signaling overhead. However, even at moderate loads our protocol offers an improvement of around 200%. The throughput comparison for the mobile network, shown in Fig. 5(c), displays similar trends, with MAC-RSV outperforming 802.11 by 70%, although the absolute throughputs for both schemes naturally decrease due to mobility. The delay comparison for the mesh networks is shown in Fig. 5(b). Lower delays for 802.11 under low loads is attributed to lower signaling overhead and inherent asynchronosity. At higher loads, the average delay of nodes using 802.11 increases sharply. On the other hand, the average delay of nodes using MAC-RSV saturates (i.e., bounded) to a nominal value. Comparison for delay in the mobile network shows a similar trend but is omitted here due to page limitations.

Finally, Fig. 5(d) compares our protocol with CATA [3]. We simulated both protocols under the same 25-node mesh network described earlier. The parameters we used for both protocols are: 0.25 persistence probability, 16 data slots and 16 reservation slots. Notice in Fig. 5(d) that throughput of CATA saturates at a moderate load of about 10Mb/s. This is because in CATA, a node can only reserve a particular data slot in the corresponding reservation slot. As a result, when the available data slots in a frame become scarce, contentions for these slots increase dramatically. On the contrary, our protocol is unaffected because nodes can contend for any data slot in any of the 16 reservation slots.

## VI. CONCLUSIONS

In this paper, we presented a synchronous and distributed medium access control protocol for multihop networks. Users contend for data slots at the beginning of a frame, and transmit data in their corresponding reserved slots. We find that resolving contentions collectively at the beginning of a frame greatly improves bandwidth utilization. Synchronization of nodes can be achieved through periodic signals from a nearby access point (as specified in IEEE 802.11 PCF) or by the use of a GPS receiver. Use of a receiver beacon by receiving nodes at the beginning of data slots protects reservations under mobility, and avoids deadlocks. Through theoretical analysis, we derive the worst-case system throughput for our protocol. Simulation results demonstrate that this new MAC far outperforms 802.11 and CATA in terms of throughput and delay even at moderate network loads. We believe that our protocol presents an attractive solution for support of real-time traffic at the MAC layer.

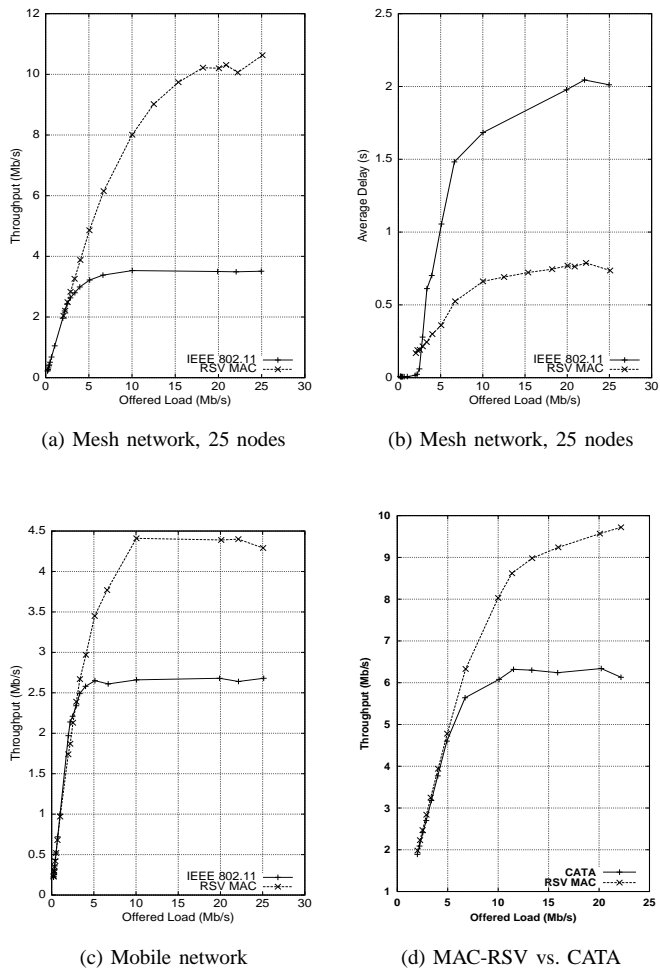

(a) Mesh network, 25 nodes  (b) Mesh network, 25 nodes

(c) Mobile network  (d) MAC-RSV vs. CATA

Fig. 5. (a)–(c): Throughput and delay comparisons of MAC-RSV with 802.11, for mesh and mobile networks. (d): Throughput comparison with CATA.